\documentclass[prl,aps,reprint,longbibliography]{revtex4-2}
\usepackage{graphicx}
\usepackage{color,soul}
\usepackage{amsmath}
\usepackage{tabularx}
\usepackage[dvipsnames]{xcolor}
\usepackage{soul}
\usepackage[normalem]{ulem}

\begin{document}

\title{The role of electron interactions in a failed insulator revealed by shot noise}

\author{Mateusz Szurek}
\affiliation{Department of Physics, Emory University, Atlanta, Georgia 30322, USA}
\email{mateusz.szurek@emory.edu}
\author{Hanqiao Cheng}
\affiliation{Department of Physics, Emory University, Atlanta, Georgia 30322, USA}
\author{Zilu Pang}
\affiliation{Department of Physics, Emory University, Atlanta, Georgia 30322, USA}
\author{Yiou Zhang}
\affiliation{Department of Physics, Emory University, Atlanta, Georgia 30322, USA}
\author{Sergei Urazhdin}
\affiliation{Department of Physics, Emory University, Atlanta, Georgia 30322, USA}


\begin{abstract}  
In materials known as failed insulators, electrical resistivity increases as temperature decreases, yet does not diverge — a phenomenon inconsistent with single-particle transport theories. We investigate the origin of this behavior by measuring shot noise in nanojunctions of nitrogen-doped $\beta$-Ta, a prototypical failed insulator and an efficient spin-orbitronic material. Junctions as short as $8$ nanometers exhibit hot-electron shot noise, indicating strong electron interactions. We show that charge hopping mediated by electron interactions can account for the anomalous failed insulator properties. Based on our findings, we identify failed insulators as a distinct class of non-Fermi liquids, opening new avenues for the studies of electron interactions and their applications.

\end{abstract}

\maketitle

\section{Introduction} 

The Mott-Anderson metal-insulator transition (MIT), arising from the interplay between disorder and electron interactions, plays a crucial role in semiconductor electronics, high-temperature superconductivity, and other exotic phenomena~\cite{Gebhard1997}. The metallic state is generally manifested by a positive temperature coefficient of resistivity (PTCR). The insulating state exhibits a negative TCR (NTCR), with the resistivity $\rho$ diverging at low $T$ due to the freeze-out of phonons mediating hopping between localized states~\cite{Liu2010}.

Non-diverging NTCR observed in thin films close to the superconductor to insulator transition was interpreted as a signature of incoherent Cooper pairing, which may suppress instead of facilitating MIT~\cite{Breznay2017,yang2019intermediate,zhang2022anomalous}. This state termed a ``weak" (or ``failed") insulator is typically associated with a large resistivity $\rho$ in the so-called Ioffe-Regel localization limit~\cite{Hussey2004}, and is sometimes described as a ``bad metal"~\cite{BadMetalTransport}. 

We define failed insulators broadly by non-diverging NTCR. This behavior is inconsistent with both diffusive transport expected for single-particle Fermi liquids and phonon-assisted hopping, and remains poorly understood. In this work, we present experimental evidence for strong electron-electron (e-e) interaction in a non-superconducting failed insulator, and show that its effects can explain the anomalous properties. We used measurements of shot noise (SN) $-$  white noise produced by biased junctions due to the discrete nature of charge carriers. The dependence of electronic noise on bias provides information about the nature of charge transport, e-e and electron-phonon (e-ph) interactions, as follows. For metallic junctions with length $L$ shorter than the e-ph scattering length $l_{e-ph}$, it can be approximated by the generalization of the dependence expected for electron diffusion~\cite{chen2023shot,Zhang2024},
\begin{equation}\label{eq:SN_general}
 S_V = 2{\frac{dV}{dI}}[eF(V_B\coth \frac{V_B}{V_{th}}-V_{th})+2k_BT],
\end{equation}
where $F$ is the Fano factor, $e$ is the electron charge magnitude, $V_B$ is the bias voltage, $dV/dI$ is the differential resistance, and $V_{th}$ accounts for thermal contribution to noise. In the analysis below, we characterize thermal effects by the broadening parameter $B=eV_{th}/2k_BT$. 
This parameter defines the crossover between thermal noise at $V_B\ll2Bk_BT$ and a linear dependence of noise on bias at $V_B\gg2Bk_BT$ characterized by the slope $dS_V/dV_B=2eFdV/dI$.

Single-electron diffusion is described by $F=1/3$, $B=1$~\cite{ShotNoiseBible}. For $l_{e-ph}\gg L\gg l_{e-e}$, where $l_{e-e}$ is the e-e scattering length, electron thermalization results in $F=\sqrt{3}/4$, $B=2/\sqrt{3}$~\cite{Yiming2024}. SN due to phonon-mediated hopping transport is also described by Eq.~(\ref{eq:SN_general})~\cite{Zhang2024}, with both parameters $F=(L_h/L)^\beta$ and $B=1/L_h$ scaled by the characteristic hopping length $L_h$. The scaling exponent $\beta=0.5-1$ of the Fano factor is determined by the Coulomb blockade effects that depend on the effective dimensionality of hopping transport~\cite{korotkov2000shot,kinkhabwala2006numerical}. In diffusive transport, SN suppression by e-ph interaction at $L\gtrsim l_{e-ph}$  results in downcurving of the bias dependence according to $S_V\propto V_B^{2/5}$~\cite{PhysRevB.49.5942}, which is often used as a signature of e-ph interaction effects~\cite{chen2023shot}. 

Recent studies~\cite{chen2023shot, PhysRevResearch.5.043143,Zhang2024,Szurek2025} suggest that these regimes break down for metals not described by the Fermi liquid theory, where e-e interactions determine electronic properties. Measurements of SN in nanowires of an archetypal ``failed" insulator $\beta$-Ta demonstrated anomalous SN suppression~\cite{Szurek2025}. Since the downcurving of $S_V(V_B)$ expected for e-ph interaction was not observed, this suppression was attributed to the presence of a correlated many-electron liquid. The Fano factors were shown to increase with decreasing nanowire length $L$, suggesting that the properties of this state may be further elucidated by studies of shorter junctions. However, planar geometry limited the accessible nanowire length to $L\geq100$~nm. 

Here, we present measurements of SN in ultrashort vertical nanojunctions of $\beta$-Ta that clarify the origin of suppressed SN and reveal a different effect of e-e interaction. We show that in short junctions the Fano factors approach the ``hot electron" values expected for efficient electron thermalization due to strong e-e interaction, which is inconsistent with the previously proposed noise suppression mechanism. We critically re-examine SN in planar nanowires and show that its suppression is dominated by the phonon relaxation into the substrate and does not directly reflect electron correlations. We develop a model that explains both the fast thermalization and the anomalous resistivity of the ``failed" insulator state by the effects of e-e interactions expected from the spin-orbit coupled electronic structure of $\beta$-Ta.  Our measurements and analysis provide insight into many-electron properties of non-Fermi liquids whose studies can elucidate exotic phenomena such as unconventional superconductivity.

\begin{figure}
\includegraphics[width=\columnwidth]{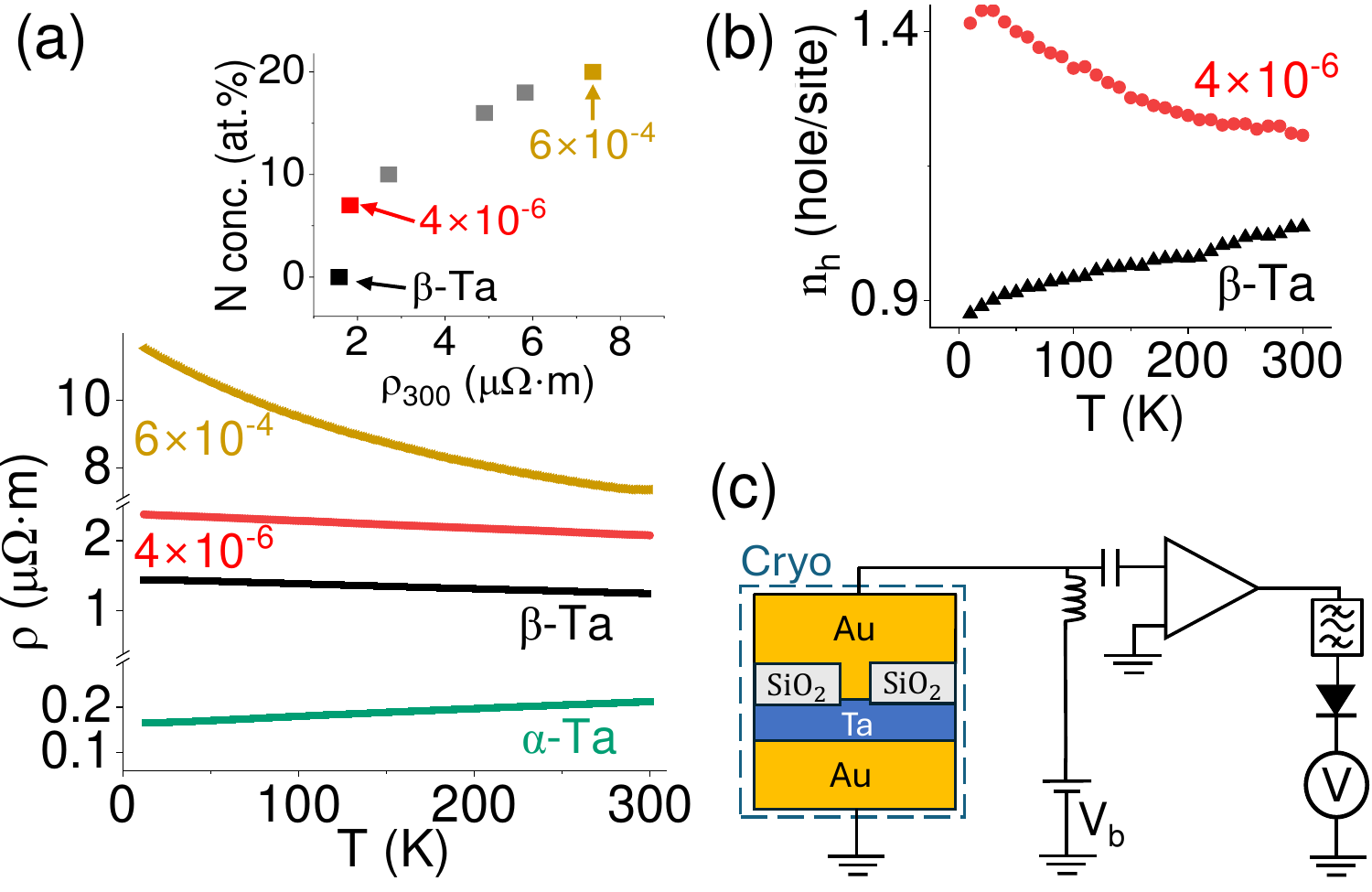}
\caption{\label{fig:1}(a) Resistivity $\rho$ vs $T$ for $\alpha$-Ta, $\beta$-Ta, and TaN$_x$ deposited at the labeled $N_2$ pressure $P_{N2}$ in Torr. Inset: nitrogen concentration for the studied films estimated from the published resistivity vs $\rho(T =300\,K)$ dependence (black and brown symbols), and measured for LD-TaN$_x$ by energy dispersive spectroscopy (red symbol) [see SM for details~\cite{supp}]. (b) Hole density per atomic site vs $T$, for $\beta$-Ta and a TaN$_x$ sample deposited at $P_{N2}=4\times10^6$~Torr. (c) Schematic of VNJ and noise measurement setup.}
\end{figure}

\section{Methods}

 We studied SN in nanojunctions based on Ta whose disordered nitrogen-doped thin films are known to exhibit a failed insulator state~\cite{Breznay2017}, and which is widely utilized as an efficient spin Hall material~\cite{liu2012spin}. Ultrashort vertical junctions (VNJs) were fabricated from multilayers deposited by ultrahigh vacuum sputtering, using a multistep e-beam lithography technique developed for nanomagnetic devices [see Supplemental material (SM) for details~\cite{supp}] ~\cite{PhysRevLett.119.257201}. The VNJ length $L$ is given by the thickness of the Ta layer partially patterned into a circular nanopillar with diameter of $40$~nm, and sandwiched between low-resistivity electrodes, Fig.~\ref{fig:1}(c).

Metallic underlayers formed by the electrodes stabilize the $\alpha$-Ta allotrope, a typical Fermi liquid [Fig.~\ref{fig:1}(a)]~\cite{Szurek2025}. To avoid this, we used a small $N_2$ pressure $P_{N2}=4\times10^{-6}$~Torr during the Ta deposition, resulting in a few percent of nitrogen incorporated in the lattice, inset in Fig.~\ref{fig:1}(a) [see SM for details~\cite{supp}]. We label this material low-doped (LD) Ta. Its resistivity is only $13\%$ higher than that of pure $\beta$-Ta, and exhibits a nearly identical temperature dependence with the ratio of resistivities varying by less than $0.5\%$ over the entire measured temperature range (Fig.~\ref{fig:1}(a)). The Hall carrier concentration is somewhat increased due to doping by nitrogen (Fig.~\ref{fig:1}(b)). Both $\beta$-Ta and LD-Ta show a modest temperature dependence of the Hall carrier concentration, with opposite trends. We tentatively attribute these effects to the partial breakdown of single-particle picture due to the large e-e interaction effects, which are shown below to depend on disorder. Nevertheless, the similarity of temperature-dependent resistivities demonstrates the same robust mechanisms of electronic transport. If the anomalous properties of $\beta$-Ta and LD Ta were associated with proximity to Mott-Anderson MIT, additional disorder introduced by nitrogen doping would result in a transition to the insulating state~\cite{PhysRevB.92.125143,Gebhard1997}. Higher nitrogen doping increases $\rho$ and enhances NTCR [Fig.~\ref{fig:1}(a)]. However, the resistivity does not diverge even for the highest doping levels within the studied range. This confirms that the failed insulator state in $\beta$-Ta cannot be explained by single-particle disorder.

SN was detected using an ultralow noise amplifier with the maximum gain of $2\times10^8$ at frequency $f_0 = 350$~MHz and $50$~MHz detection bandwidth. Its output was bandpass filtered and converted into dc voltage by a microwave rectifier, Fig.~\ref{fig:1}(c). Noise detection in the microwave range provided a high sensitivity while avoiding the low-frequency EMI and flicker noise. Details of measurement approach and calibration are discussed elsewhere ~\cite{Szurek2025} [also see SM~\cite{supp}].

\begin{figure}
\includegraphics[width=0.9\columnwidth]{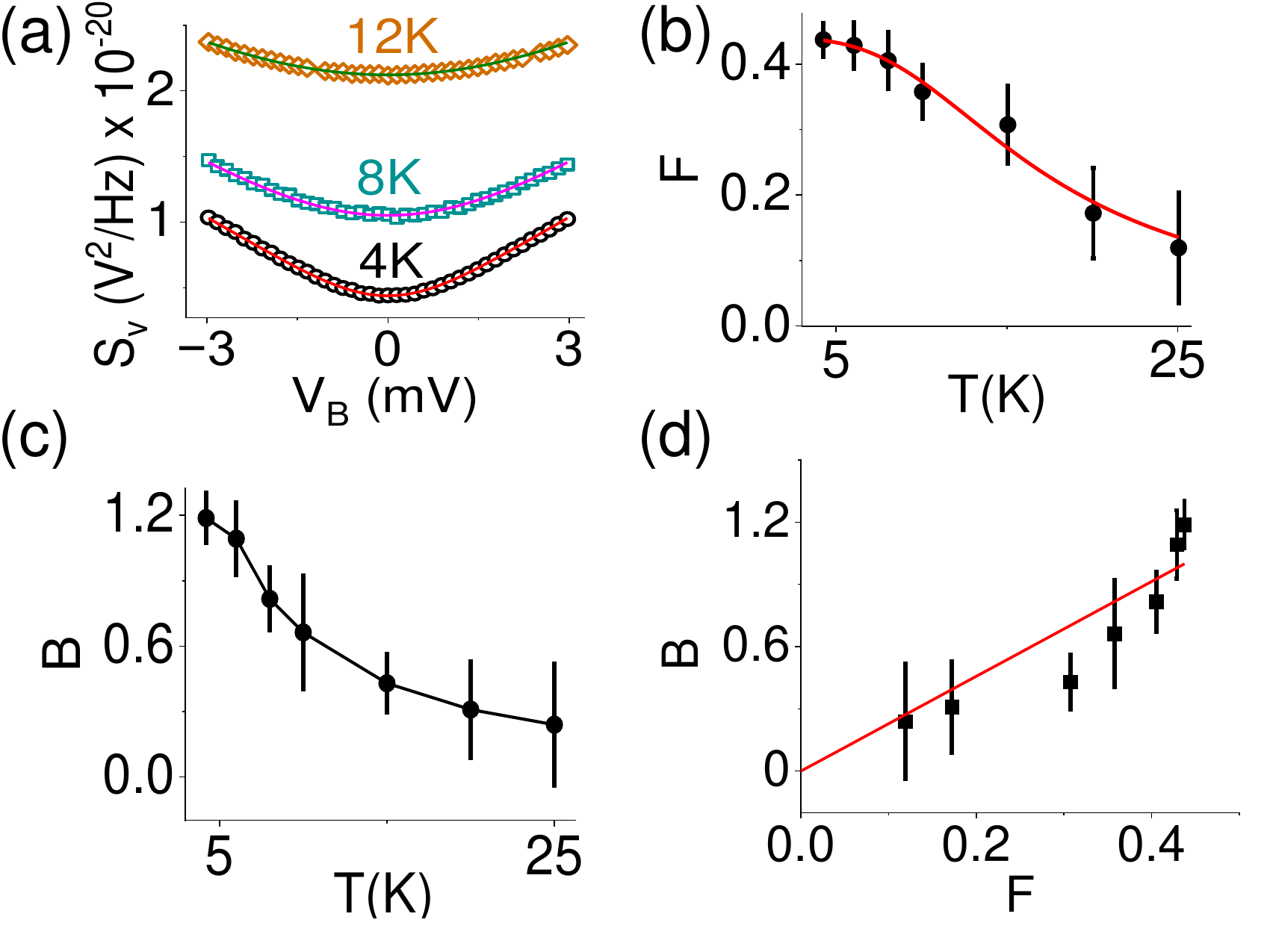}
\caption{\label{fig:2} Noise measurement results for the LD-Ta VNJ with $L = 8$~nm. (a) Noise vs $V_B$, at the labeled $T$. Curves: fittings with Eq.~(\ref{eq:SN_general}). (b) $F$ vs $T$. Curve: fitting as described in the text. (c) $B$ vs $T$. Line is a guide for the eye. (d) $B$ vs $F$. The line is a linear fit with zero intercept.}
\end{figure}

\section{Results}

 Examples of the bias-dependent noise are shown in Fig.~\ref{fig:2}(a) for an LD-Ta VNJ with $L=8$~nm. The dependence is linear at large $V_B$ and saturates at small $V_B$, which is well-approximated by Eq.~(\ref{eq:SN_general}) [solid curves in Fig.~\ref{fig:2}(a)]. The data do not exhibit any downcurving expected for the effects of e-ph interaction on noise~\cite{SNDiffusive,Henny1997}. The temperature-dependent Fano factors and thermal broadening determined from the fitting are shown in Figs.~\ref{fig:2}(b),(c). At $T=4.2$~K, they are close to $\sqrt{3}/4\approx0.43$ and $2/\sqrt{3}\approx1.15$, respectively, expected for the hot-electron regime, suggesting a thermalized electron state due to strong e-e interactions. We note that noise produced by thermalized electrons is determined by the thermodynamics of electron system, which is expected to result in a linear dependence on bias at $V>V_{th}$ regardless of the interaction details.

The Fano factor decreases with increasing $T$, Fig.~\ref{fig:2}(b). A similar suppression in nanowires of $\beta$-Ta~\cite{Szurek2025} and heavy fermion ``strange" metals~\cite{chen2023shot} was attributed to a correlated electron liquid. However, if electrons are thermalized at $T=4.2$~K, they are also likely thermalized at higher $T$, so noise contains information only about the effective electron temperature $T_e$. In this case, the decrease of $F$ indicates that $T_e$ is reduced by an additional dissipation channel, which can be described by the diffusion equation~\cite{PhysRevB.49.5942,Henny1997}
\begin{equation}\label{eq:diffusion}
\sigma\left(\frac{ V_B}{L}\right)^2=-\frac{\partial}{\partial x}\left(\frac{\kappa_e\partial T_e}{\partial x}\right)+f,    
\end{equation}
where $\sigma$ is conductivity, $\kappa_e$ is the electron thermal conductivity, and $f(T_e,T)$ is the rate of dissipation into the additional channel. For negligible $f$, Wiedemann-Franz law $\kappa_e=\sigma T L_0$ leads to a linear dependence of noise on bias at $V_B>V_{th}$. Here, $L_0$ is the Lorenz number. A finite $f$ typically leads to a nonlinear dependence. For instance, for e-ph coupling described by $f\propto(T^5_e-T^5)$, one obtains $S_V\propto V_B^{2/5}$ at $T_e\gg T$~\cite{SNDiffusive}. To reproduce the linear $S_V(V_B)$ observed at $V>V_{th}$, one can empirically approximate $f$ by the same form as the electron contribution, $f(T_e,T)=-\frac{f_0}{2}\partial^2 T^2_e/\partial x^2$, where $f_0$ is a function of only temperature $T$, giving
\begin{equation}\label{eq:fano_diss}
F(T)=\frac{\pi k_B}{4e\sqrt{L_0+f_0(T)}}.
\end{equation}

This mechanism describes suppressed $F$, but also increased thermal broadening, $B=2e\sqrt{L_0+f_0}/\pi k_B$. Instead, we find that $B$ is approximately proportional to $F$, Fig.~\ref{fig:2}(d). This indicates that the dependence on bias is not fully captured by the proposed empirical form of $f$. We note that the linear dependence with suppressed Fano factor observed at higher bias cannot be described by the models based on the Fermi liquid picture~\cite{SNDiffusive}. Anomalous thermal broadening may thus reflect unconventional dissipation mechanisms in the non-Fermi liquid electron state, which is empirically parametrized by Eq.~(\ref{eq:fano_diss}) and analyzed in detail below.

\begin{figure}
\includegraphics[width=\columnwidth]{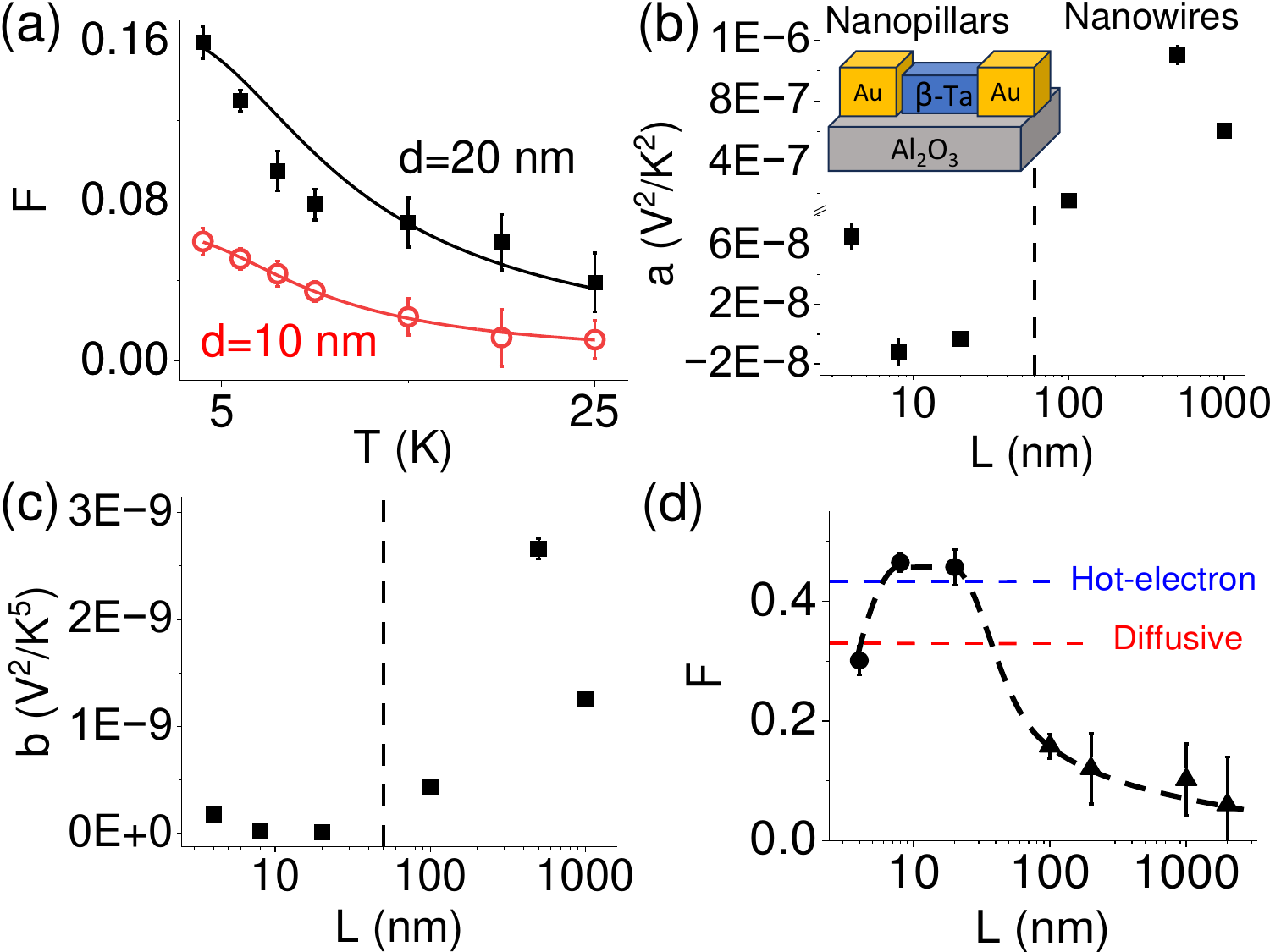}
\caption{\label{fig:3} (a) $F$ vs $T$ for two $\beta$-Ta nanowires with the same length $L=500$ nm and cross-section areas, but different contact area with the substrate as determined by thickness. Curves: fittings as described in the text. (b), (c)  Fitting parameters in the dependence $f_0=a+bT^3$ vs $L$, for VNJs and nanowires as labeled. Lines are guides for the eye. To maximize the precision of $a$, fitting was performed for $T\leq10$~K. Inset in (b): nanowire schematic. (d) $F$ vs $L$ at $T = 4.2$~K for LD-Ta VNJs ($L\leq20$~nm) and $\beta$-Ta nanowires ($L\geq 100$ nm).}
\end{figure}

The dissipation leading to suppressed $F$ and described by $f(T,T_e)$ can be mediated by phonons or electronic degrees of freedom such as spin or collective excitations that do not contribute to charge transport~\cite{chen2023shot}. To elucidate the possible phonon contribution, we also studied nanowires of pure $\beta$-Ta deposited directly on sapphire substrates and patterned by e-beam lithography [inset in Fig.~\ref{fig:3}(b)]. Similarly to VNJs, the dependence of noise on bias is well-approximated by Eq.~(\ref{eq:SN_general}) [see SM~\cite{supp}]. Figure~\ref{fig:3}(a) shows $F(T)$ for two nanowires with the same length $L=500$~nm and cross-section areas but different aspect ratios: thickness $d=10$~nm, width $w=750$~nm for one nanowire, and $d=20$~nm, $w=375$~nm for the other. Both electrical and thermal transport characteristics of the two nanowires themselves were similar, but the contact areas with the substrate differed by a factor of $2$. For the wire with the larger contact area, $F$ is significantly smaller at $4.2$~K and more rapidly decreases with decreasing $T$, as expected for the phonon effects due to more efficient dissipation into the substrate. We note that multi-phonon effects may influence the temperature dependence of hopping rates but would still  result in enhancement of thermal broadening that scales with the nanowire length~{\cite{Banerjee2016}}. However, none of the studied nanowires showed an enhancement of thermal broadening parameter $B$.

The function $f_0(T)$ obtained from the temperature dependence of the Fano factor is well approximated by the empirical relation $f_0(T)=a+bT^3$ for all the studied structures, as illustrated in Fig.~\ref{fig:2}(a) for a VNJ, and in Fig.~\ref{fig:3}(a) for two nanowires with thicknesses $d=10$ and $20$~nm. We note that the $20$~nm the data fall below this dependence at $T=6-8$~K, while the $T=20$~K point is higher. The contact area of this nanowire with the substrate is smaller than that of the $10$~nm nanowire. Consequently, these anomalies may reflect sensitivity to interface imperfections affecting the temperature dependence of the dissipation into the substrate.

 A finite $a>0$ reflects suppression of $F$ due to the additional dissipation in the limit $T\to0$, while $b>0$ describes its enhancement with increasing $T$. Both $a$ and $b$ are significantly larger for nanowires than VNJs, consistent with additional dissipation into the substrate. Two important conclusions follow from these dependences. First, a large value of $a$ for nanowires implies that SN is suppressed even in the limit $T\to0$, contrary to the expectation for weak e-ph coupling in Fermi liquids~\cite{SNDiffusive,stierbach1996hotElectron}. Second, the small value of $b$ in VNJs shows that thermal effects on $F$ are small at $T=4.2$~K. For instance, for the $L=8$~nm VNJ, $b=1.3\times10-11\,V^2/K^5$ gives $F(T=0)$ only $4\%$ larger than $F(4.2\,K)$. We note that fitting gives a small $a<0$ for both $L=8$~nm and $20$ nm, signifying that at $T\to0$ SN may be slightly enhanced compared to the hot-electron regime. This may be an indication of the electron correlation effects, but the difference is too small for a definitive conclusion. We also note that for $L = 4$ nm, both $a$ and $b$ substantially deviate from the values for larger $L$, indicating a change in the transport regime at ultrashort length scales.

\begin{figure}
\includegraphics[width=\columnwidth]{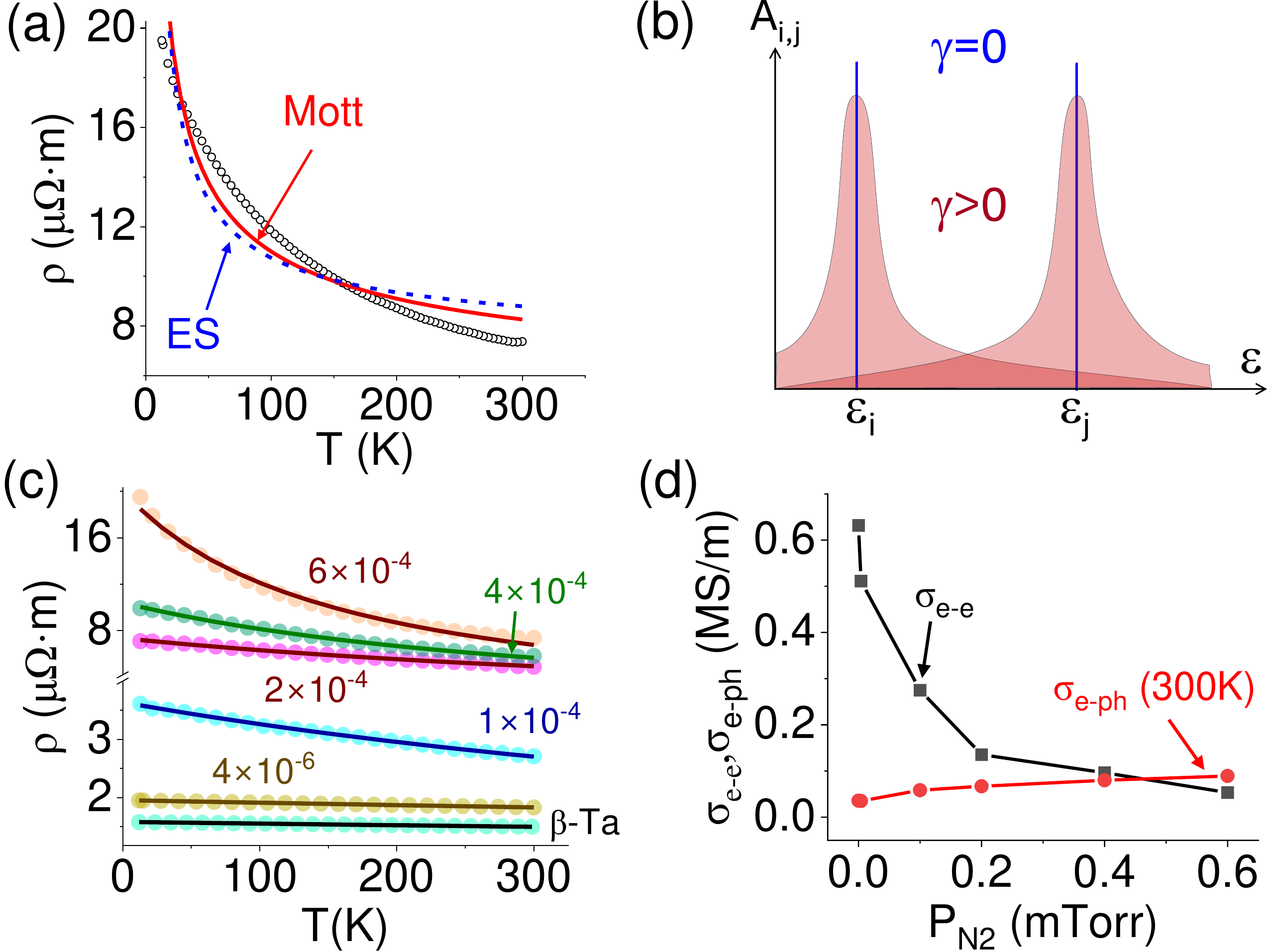}
\caption{\label{fig:4}  
(a) $\rho$ vs $T$ for HD-Ta and  best fits with Mott and ES VRH models, as labeled. (b) Spectral density functions of two quasi-localized states for negligible lifetime broadening ($\gamma = 0$) and finite lifetime due to e-e interaction ($\gamma>0$). (c) $\rho$ vs $T$ for $\beta$-Ta and a series of nitrogen-doped Ta films, with fittings using the model of hopping mediated by the e-e interaction, as discussed in the text. (d) $\sigma_{e-e}$ and $\sigma_{e-ph}$ vs nitrogen pressure at $T = 300$~K.}
\end{figure}

Based on these observations, we can infer the relevant transport regimes from the noise measurements at $T=4.2$~K, Fig.~\ref{fig:4}(a). The Fano factor for VNJs with $L=8$~nm and $20$~nm is close to $F=\sqrt{3}/4$ expected for thermalized electrons. For $L=4$~nm, it approaches $F=1/3$ expected for single-electron diffusion, suggesting that $4\,nm\lesssim l_{e-e}\lesssim 8$~nm, $l_{e-ph}>20$~nm at $4.2$~K. Finally, the Fano factor is suppressed in nanowires due to e-ph interaction. The thermal broadening was small in all the samples, $B<1/F$, eliminating phonon-mediated hopping as the transport mechanism [see SM for details~\cite{supp}]. The dependence $F(L)$ is similar to that in noble metal wires~\cite{stierbach1996hotElectron}, but with at least three orders of magnitude smaller characteristic lengths. 

\section{Analysis and Discussion}

The central question raised by our measurements is whether electrons form a Fermi liquid of extended Bloch waves, or hop among states that do not carry current. The Fermi liquid picture is meaningful only if the scattering length of electron waves significantly exceeds lattice spacing. Using the measured resistivity $\rho$ and the Hall coefficient [Fig.~\ref{fig:1}, see also SM~\cite{supp} for ellipsometry], we obtain the effective Drude mean free path (mfp) of $\beta$-Ta and LD-Ta $l\lesssim~0.3$~nm close to the inter-atomic spacing $a=0.28$~nm. This unphysically small value indicates that the Drude-Sommerfeld model is inapplicable. The observed NTCR is also inconsistent with the Fermi liquid picture, where a PTCR would be expected due to temperature-dependent phonon scattering. Furthermore, a short thermalization length inferred from our SN measurements places a strict limit on the extent of phase-coherent single-electron states. 

The limit $l\lesssim a$ is commonly interpreted as a criterion for the Mott-Anderson localization~\cite{Hussey2004}, which is qualitatively consistent with NTCR due to the phonon-mediated variable-range hopping (VRH) transport among localized states~\cite{Gebhard1997,Liu2010}. To assess this possibility, we fit the resistivities with Mott and Efros-Shklovskii (ES) VRH models, $\rho(T) =  \rho_0 \exp\left[ \left( T_0/T\right)^d \right]$, where $d=1/4$ for Mott's VRH and $1/2$ for ES hopping, and $\rho_0$, $T_0$ are determined by the electron density and disorder. Neither model fits the $\rho(T)$ data for the studied films, as illustrated in Fig.{~\ref{fig:4}}(a) for one the films. Single-electron hopping is also inconsistent with the lack of enhanced thermal broadening in our SN measurements. 

We now discuss the assumptions underlying these models, analyze their relevance to $\beta$-Ta, and introduce an extension that provides a consistent interpretation of our results. VRH describes single-electron hopping among single-particle states with well-defined energies $\epsilon_i$, $\epsilon_j$. Energy conservation requires that a phonon with energy $\hbar\omega=|\epsilon_i-\epsilon_j|$ is absorbed or emitted in this process. We contrast this behavior with the tight-binding models of a Fermi liquid, which also describe single-electron hopping between atomic orbitals with different energies. In contrast to VRH, hybridization between wavefunction amplitudes on different sites results in phase-coherent extended states. These models are distinguished by the effects of wavefunction coherence. Electron interaction with the phonon bath decoheres the wavefunction phase at a rate $\gamma$. If dephasing is faster than the hopping rate, $\gamma> V_{ij}$, the dynamics can be approximated as classical particle motion, which is the basis of VRH models. Below, we show that the condition $\gamma>V_{ij}$ can be satisfied in $\beta$-Ta due to the small interatomic hopping amplitudes and fast dephasing by e-e interaction. 

\textit{Suppression of electron hopping in $\beta$-Ta.} In a simple metal with valence charge density of about one hole per atom [Fig.~\ref{fig:1}(b)], elementary charge transport  events involve hopping between neighboring sites instead of distant impurity levels. Remarkably, we find that the interatomic hopping matrix elements among the 5d valence states in $\beta$-Ta are anomalously small. Each atom in $\beta$-Ta is coordinated with $20$ nearest and next-nearest neighbors within $1.4\%$ distance~\cite{Jiang2003-ye}. This coordination produces an almost spherically symmetric effective crystal field whose effects can be neglected in the lowest-order approximation with respect to the fine-structure splitting by spin-orbit coupling (SOC). In $\beta$-Ta, the $5d$ levels are split SOC into the high-energy $J=5/2$ sextuplet and the low-energy $J=3/2$ quadruplet. The $J=5/2$ sextuplet is unfilled, while the quadruplet is filled with the three valence electrons leaving one $J=3/2$ valence hole. This analysis agrees with the valence charge density of 1 hole per site determined from the Hall measurements [Fig.~\ref{fig:1}]. The spin-orbit structure of the $J=3/2$ states is
\begin{equation}\label{eq:spin-orbit}
	\begin{cases}
		|m_j = \frac{3}{2} \rangle = -\sqrt{\frac{4}{5}} |2,\downarrow \rangle + \sqrt{\frac{1}{5}} |1,\uparrow \rangle\\
		|m_j = \frac{1}{2} \rangle = -\sqrt{\frac{3}{5}} |1,\downarrow \rangle + \sqrt{\frac{2}{5}} |0,\uparrow \rangle\\
		|m_j = -\frac{1}{2} \rangle = -\sqrt{\frac{2}{5}} |0,\downarrow \rangle + \sqrt{\frac{3}{5}} |1,\uparrow \rangle\\
		|m_j = -\frac{3}{2} \rangle = -\sqrt{\frac{1}{5}} |-1,\downarrow \rangle + \sqrt{\frac{4}{5}} |-2,\uparrow \rangle,
	\end{cases}
\end{equation}
where the vector $|m_l,m_s \rangle$ denotes orbital and spin projections on the selected quantization axis. 

We use the Koster-Slater method to determine the hopping matrix elements $\langle m_j | m_j',\vec{r} \rangle$ between a Ta atom and its neighbor at the position $\vec{r} = (r,\theta,\phi)$. Here, $\theta$ is the angle between $\vec{r}$ and the quantization axis, and $\phi$ is the polar angle. The hopping amplitudes between spin-orbit coupled states Eq.~(\ref{eq:spin-orbit}) can be expressed in terms of the Koster-Slater integrals $V_{dd\sigma}$, $V_{dd\pi}$, and $V_{dd\delta}$~\cite{slater1954simplified}. As shown in Table~\ref{tbl:J_hopping}, they are determined by just two combinations $V_1 = V_{dd\sigma}+V_{dd\pi}-2V_{dd\delta}$ and $V_2 = 2V_{dd\pi}+8V_{dd\delta}$. In the atomic-sphere approximation (ASA), the Koster-Slater integrals are related by $V_{dd\sigma} = 6V_{dd\delta} = -3V_{dd\pi}/2$~\cite{jenke2021tight}, giving $V_1=V_2=0$ and thus vanishing hopping amplitudes. The ASA approximation is violated by the effects of crystal field on the atomic orbital symmetry. Nevertheless, this analysis shows that the hopping amplitudes are anomalously small. Importantly, hopping suppression originates from the small hybridization between atomic orbitals mixed by the spin-orbit interaction, which is a property of the  $\beta$-Ta crystal structure that does not require disorder. This result explains why disorder introduced by nitrogen doping has little effect on the electronic properties of $\beta$-Ta.

\begin{table*}
	\caption{Hopping amplitudes between $J=3/2$ states}
	\label{tbl:J_hopping}
	\begin{tabular}{ccccc}
		\hline
		& $| \frac{3}{2},\vec{r} \rangle$ & $| \frac{1}{2},\vec{r} \rangle$ & $| -\frac{1}{2},\vec{r} \rangle$ & $| -\frac{3}{2},\vec{r} \rangle$\\
		\hline
		\hline
		$\langle \frac{3}{2} |$   & $\frac{1}{10}(3V_1 \sin^2\theta + V_2)$    & $-\frac{\sqrt{3}}{10} V_1 \sin2\theta e^{-i\phi}$ &   $-\frac{\sqrt{3}}{10} V_1 \sin^2\theta e^{-2i\phi}$ & $0$\\
		$\langle \frac{1}{2} |$   & $\frac{\sqrt{3}}{10} V_1 \sin2\theta e^{i\phi}$    & $\frac{1}{10}(3V_1 \cos^2\theta + V_1 + V_2)$ &   $0$ & $-\frac{\sqrt{3}}{10} V_1 \sin^2\theta e^{-2i\phi}$\\
		$\langle -\frac{1}{2} |$   & $-\frac{\sqrt{3}}{10} V_1 \sin^2\theta e^{2i\phi}$    & $0$ &   $\frac{1}{10}(3V_1 \cos^2\theta + V_1 + V_2)$ & $\frac{\sqrt{3}}{10} V_1 \sin2\theta e^{-i\phi}$\\
		$\langle -\frac{3}{2} |$   & $0$    & $-\frac{\sqrt{3}}{10} V_1 \sin^2\theta e^{2i\phi}$ &   $-\frac{\sqrt{3}}{10} V_1 \sin2\theta e^{i\phi}$ & $\frac{1}{10}(3V_1 \sin^2\theta + V_2)$\\
		\hline
	\end{tabular}
\end{table*}

\textit{Dephasing due to e-e interaction.} The dephasing required for the hopping transport originates in VRH from electron interaction with the thermal phonon bath. Since temperature has little effect on the electronic properties of $\beta$-Ta, single-electron dephasing must originate from the properties of the electronic state unrelated to phonons. We now show that e-e interaction provides such a mechanism, which is much more efficient than in semiconductors due to the high carrier density.

To elucidate this mechanism, we use a minimal two-site model. Choosing the z-axis along the direction between the sites, the hopping matrix given by Table~\ref{tbl:J_hopping} becomes diagonal, which reflects the conservation of the projection $m$ of angular momentum on the cylindrical symmetry axis. We consider two spin-orbit coupled atomic states $m_1$, $m_2$ that form a pseudospin $\sigma$. The Mott Hamiltonian on these states is 
\begin{equation}\label{eq:Mott-Hubbard}
H=-V\sum_\sigma (c^\dagger_{1\sigma}c_{2\sigma}+c^\dagger_{2\sigma}c_{1\sigma})+U\sum_in_{i\uparrow} n_{i\downarrow},
\end{equation}
where $U$ is the effective onsite Mott interaction, $c_{i\sigma}$ are particle operators, and $V$ is the hopping amplitude whose dependence on the pseudospin is neglected. 

Assume that a pseudospin-up hole is initially in a phase-coherent state with relative phase difference $\phi=0$ between the two sites. The fluctuating density $\delta n_i(t)$ in the pseudospin-down channel modifies the site energies of the pseudospin-up hole via Mott interaction according to $\delta\epsilon_{i\uparrow}=U\delta n_{i\downarrow(t)}$, resulting in the relative phase
\begin{equation}\label{eq:phase}
\phi(t)=\frac{U}{\hbar}\int_0^t[\delta n_1(t')-\delta n_2(t')]dt'.
\end{equation}
Since many electrons hop through each site, we assume that the fluctuations are uncorrelated between the two sites, and their time dependence can be approximated by a Markovian diffusion process with the hopping rate $V/\hbar$, giving~\cite{RevModPhys.75.715}
\begin{equation}\label{eq:population}
\langle\delta n_i(t_1)\delta n_i(t_2)\rangle=\langle\delta n^2_i(0)\rangle e^{-V|t_1-t_2|/\hbar}.
\end{equation}
Combining Eqs.(\ref{eq:phase}), (\ref{eq:population}), we obtain
\begin{equation}
\begin{split}
\langle\phi^2(t)\rangle=\frac{U^2}{\hbar^2}\int_0^t\int_0^t\delta n_i(t_1)\delta n_i(t_2)dt_1dt_2\\
\approx2U^2t\langle\delta n_i^2\rangle/\hbar V,
\end{split}
\end{equation}
where the last equality is obtained in the approximation $t\gg \hbar/V$. The corresponding dephasing rate due to e-e interaction is $\gamma_e=2\langle\delta n_i^i\rangle U^2/\hbar V$. In the hopping regime, it should be larger than the hopping rate $V/\hbar$, which requires $V<\sqrt{2\langle\delta n_i^i\rangle}U$. The variance of density can be estimated by assuming that the total average site population $n_{tot}=1$ is randomly distributed over the $J=3/2$ quadruplet so each channel has average population $\nu=1/4$, giving $\langle\delta n_i^2\rangle=\nu(1-\nu)=3/16$. This analysis shows that the hopping regime in $\beta$-Ta results from suppressed $V$ and large charge fluctuations due to the partial filling of atomic levels. We identify this mechanism with localization, in the sense that the spatial extent of phase-coherent single-electron wavefunction is limited to atomic scale due to its efficient dephasing by e-e interaction. In contrast to the Mott-Anderson mechanism, MIT is avoided because the hole dynamics in the $J=3/2$ quadruplet filled with 1 hole is not constrained by the Pauli principle.

\textit{Hopping mediated by e-e interactions.} In the VRH models, phonons provide the activation energy $\Delta\epsilon=\epsilon_i-\epsilon_j$ for hopping between different impurity levels $\epsilon_i$, $\epsilon_j$. We show that e-e interaction provides an activation mechanism between single-particle levels $\epsilon_i$, $\epsilon_j$ similar to the effect of e-ph interaction. Since the projection $m$ of angular momentum on the hopping direction is conserved, one may naively assume that in the absence of disorder $\epsilon_i=\epsilon_j$. However, the unit cell of $\beta$-Ta contains 30 atoms with non-equivalent neighboring sites, resulting in a finite $|\epsilon_i-\epsilon_j|\sim V$ even in the absence of disorder. 

Based on the Mott-Hubbard model developed above, the single-particle atomic levels fluctuate due to the charge density fluctuations. In the two-state pseudospin approximation introduced above, $\langle\delta\epsilon^2\rangle=U^2\langle\delta n^2\rangle\approx3U^2/16$. Since the ground state energy of the many-electron system is well-defined, this fluctuation represents the uncertainty of single-particle energy due to its entanglement with other electrons, and based on the uncertainty principle can be also interpreted as a finite lifetime $\tau_{e-e}=h/\gamma$ of single-particle states, where $\gamma=\sqrt{\delta \epsilon^2}$. Since $\gamma$ represents the rate of energy exchange between a single electron state and other electrons, it is also the thermalization rate in the non-equilibrium state. The short electron thermalization length $l_{th}<8$~nm observed in our SN measurements is then explained by the effects of large charge fluctuations $\langle\delta n^2\rangle$ resulting in efficient energy exchange between localized (i.e. dephased) holes.

The effects of single-particle energy uncertainty on hopping can be analyzed using an extension of the Fermi golden rule for the transition rate between sites $i$ and $j$ [Eq.(8.69) in Ref.~{\cite{bruus2004}],
\begin{equation}\label{eq:rate}
\Gamma_{e-e,i\to j} = \frac{2\pi}{\hbar} n_i(1-n_j)|t_{ij}|^2\int_{E}A_i(E)A_j(E)dE,
\end{equation}
where $n_{i,j}$ are populations, and $A_i(E)=-\frac{1}{\pi}Im G_i(E)$ is the spectral function describing a single-particle level. Here, $G_i(E) = (E - \epsilon'_i -\Sigma)^{-1}$ is the Green's function,  $\Sigma=\Sigma'-i\Sigma"$ is the self-energy determined by e-e interactions. Its real part determines the dressed single-particle energy $\epsilon(i)=\epsilon'_i-\Sigma'$. In the Mott-Hubbard model described above, $\Sigma'=U\langle n_\uparrow n_\downarrow \rangle$, which generally includes the single-particle mean-field Coulomb repulsion, as well as correlations between single-particle populations due to entanglement. Meanwhile, the imaginary part $\Sigma"=\gamma$ is given by the single-particle energy broadening.

In the limit of negligible broadening $\gamma\to0$, the single-particle energy is well-defined, $A_i(E)\to\delta(E-\epsilon_i)$, Fig.~\ref{fig:4}(b). Hopping between states with energies $\epsilon_i\ne\epsilon_j$ described by Eq.~(\ref{eq:rate}) is prohibited, consistent with the zero-temperature limit of single-electron hopping models. However, at $\gamma>0$ the spectral functions $A_{i,j}(E)=\gamma/[(E-\epsilon_{i,j})^2+\gamma^2]$ overlap, enabling hopping even in the absence of phonons.

To describe the effects of e-e interaction-mediated hopping on conductivity, we propose a simple empirical model in the spirit of the parallel resistor formula (PRF) commonly applied to metals with PTCR in the Ioffe-Regel limit~\cite{Hussey2004}. Our approximation extends this model to NTCR. We assume that e-e interaction-mediated hopping provides a transport channel parallel to phonon-mediated hopping, which is dominant below the VRH localization temperature $T_0$. At $T>T_0$, both Mott and ES VRH models give $\rho_{e-ph}(T)=\rho_{e-ph}(T_0)T_0/T$ reflecting the Raleigh-Jeans law for the phonon population. On the other hand,  the rate of e-e interaction-mediated hopping is temperature independent if disorder and lifetime broadening dominate over thermal effects. The formula $\frac{1}{\rho(T)}=\frac{1}{\rho_{e-e}}+\frac{T}{T_0\rho_{e-ph}(T_0)}$ provides a good approximation for the temperature-dependent resistivities of $\beta$-Ta and all nitrogen-doped Ta films, Fig.~\ref{fig:4}(c). 

According to our model, nearly constant resistivity of pure $\beta$-Ta indicates that the contribution of hopping mediated by e-e interaction to the conductivity is dominant. In contrast, the two contributions to hopping become comparable at high temperatures and large nitrogen doping. These results are consistent with our Mott-Hubbard model: in the limit of purely disorder-driven localization, charge fluctuations vanish so the effects of e-e interaction on both hopping and localization are reduced. In Fig.~\ref{fig:4}(d), this is reflected by the rapid reduction of the conductivity $\sigma_{e-e}$ with increased doping. This competition between interactions and disorder is distinct from the Mott-Anderson MIT mechanism.

Localization driven by e-e interaction can be interpreted as a consequence of dressing, qualitatively similar to the polaron formation due to e-ph interactions. In weakly-interacting ordered systems, e-e interactions renormalize the band properties of electrons, similar to the mobile Fröhlich polaron. Quasi-particle lifetime diverges at vanishing energy $E\to0$ relative to the Fermi surface due to the constraints imposed by momentum conservation, and the corresponding self-energy is real. Our Mott-Hubbard model demonstrates that in systems that satisfy the Mott localization criterion $U\gtrsim V$ but avoid Mott-Anderson localization due to the their multi-orbital valence structure, e-e interactions can result in small polaron-like states localized by single-particle dephasing. Their dynamics is not constrained by the single-particle momentum or energy conservation due to the exchange with the electron bath. Thus, ``failed" insulators realize an extreme limit of energy-dependent single-particle relaxation, the other extreme being Fermi liquids (Ohmic relaxation $\tau\propto 1/E^2$), with ``strange" metals characterized by Planckian relaxation ($\tau\propto1/E$) representing an intermediate case~\cite{RevModPhys.94.041002}.

\section{Summary}

Our shot noise measurements in ultrashort nanojunctions of nitrogen-doped Ta revealed electron thermalization over the length not exceeding $8$~nm at $4.2$~K, indicating large effects of electron-electron interactions. Analysis shows that these interactions result in dephasing leading to localization. Localization and MIT are usually interpreted as synonymous. 
In contrast, we showed that e-e interactions can also mediate charge hopping between single-particle states, preventing MIT. This mechanism is analogous to the effects of electron-phonon interaction in phonon-mediated hopping but is independent of temperature, providing an explanation for the `failed" insulator behaviors characterized by non-divergent resistivity. 

In spin-orbitronics, $\beta$-Ta is known as one of the most efficient spin Hall effect (SHE) sources, which has been traditionally attributed to spin-orbit-coupled single-electron band structure~\cite{liu2012spin}. The effects of localization and strong interactions demonstrated by our study are unexplored in the context of SHE. Intriguingly, alloying of conventional Fermi liquids results in a relatively abrupt and significant enhancement of SHE on approaching the Ioffe-Regel regime corresponding to the onset of the ``failed insulator" state~\cite{Shashank2025,liu2025enhancingzspingeneration}. Qualitatively, we expected that a large SHE results from the suppressed quenching of the spin-orbit structure of atomic states by the crystal field. This observation can provide guidance in the ongiong search for efficient SHE materials.

Superconductivity observed in many ``failed" insulators~\cite{Breznay2017,yang2019intermediate,zhang2022anomalous} is commonly attributed to the conventional mechanism in the ``dirty" limit. Our analysis shows that the single-particle properties of these materials are strongly influenced by the Mott interaction. This connection with unconventional superconductivity warrants further studies. New insights may be also provided by the theory of conventional phonon-mediated superconductivity that explains the mechanism of long-range many-particle coherence among localized single-particle states, similar to superfluidity in He$^3$.

We acknowledge support from the SEED award from the Research Corporation for Science Advancement and the NSF award ECCS-2448290. Shot noise analysis by Y.Z was supported by the subcontract No.C5808 for the US Department of Energy award No.88148 through Los Alamos National Laboratory.




\bibliography{bTa}

\end{document}